\documentclass[epj]{webofc}
\usepackage[varg]{txfonts}   % Web of Conferences font
\woctitle{MESON2014 - the 13$^\textrm{th}$ International Workshop on Meson Production, Properties and Interaction}

\newcommand{\ten}[1]{$\ 10^{-#1}$}

\begin{document}
\selectlanguage{english}
\title{$B_s$ mesons: semileptonic and nonleptonic decays}

% insert email only for speaker/presenter
\author{C. Albertus \inst{1}\fnsep\thanks{\email{albertus@ugr.es}} \and
        E. Hern\'andez\inst{2} \and
        C. Hidalgo-Duque\inst{3} \and
        J. Nieves\inst{3}
% comment out the next line if not needed
%       \\for the XXXXX Collaboration
}

\institute{ Departamento de F\'\i sica At\'omica, Nuclear y Molecular 
e Instituto Carlos I de F\'\i sica Te\'orica
y Computacional, Universidad de Granada, Avenida de Fuentenueva s/n, E-18071 Granada, Spain
\and
           Departamento de F\'\i sica Fundamental e IUFFyM, Universidad de Salamanca, 
Plaza de la Merced s/n, E-37008 Salamanca, Spain
\and
           Instituto de F\'\i sica Corpuscular (IFIC), Centro Mixto CSIC -Universidad de Valencia,
Institutos de Investigaci\'on de Paterna, Apartado 22085, E-46071 Valencia, Spain
          }

\abstract{% 
  In this contribution we compute some nonleptonic and
  semileptonic decay widths of $B_s$ mesons, working in the context of
  constituent quark models \cite{Albertus:2014gba, Albertus:2014bfa}. For the case of
  semileptonic decays we consider reactions leading to kaons or
  different $J^\pi$ $D_s$ mesons. The
  study of nonleptonic decays has been done in the factorisation
  approximation and includes the final 
  states enclosed in Table~\ref{tab:brcomp2}.
}
\maketitle
\vspace{-0.5cm}
\section{Introduction}
The description of CP violation in the Standard Model demands an
accurate knowledge of the Cabibbo-Kobayashi-Maskawa (CKM) quark mixing
matrix. Being the b-sector the one known with lesser precision,
a precise quantitative study of the weak decays of $B$ and $B_s$
mesons is needed. In this contribution we summarise our studies of the
semileptonic decays $B_s$ into $K$ and $D_s$ estates, and some of
the nonleptonic decays studied in  
Refs.\cite{Albertus:2014gba, Albertus:2014bfa}, evaluated
in the context of a constituent quark model.
\section{Semileptonic $B_s\to K$ decay}
The hadronic matrix element for this decay can be parametrised in
terms of the form factors $f_+$ and $f_0$. If we neglect the mass of
the leptons, only $f_+$ contributes to the differential decay width
\begin{eqnarray}
\frac{d\Gamma}{dq^2}=\frac{G_F^2}{192\pi^3}\,|V_{ub}|^2\,
\frac{\lambda^{3/2}(q^2,M_{B_s}^2,M_K^2)}{M_{B_s}^3}\,f_+^2(q^2)
\label{eq:dgdq2}
\end{eqnarray}
with $G_F$ being the Fermi constant, $|V_{ub}|$ the modulus of the
corresponding CKM matrix element and $\lambda(a,b,c)=a^2
+b^2+c^2-2ab-2ac-2bc$.

We evaluate the valence quark contribution to the form factor that we supplement
with a $B^*$-pole one to improve its behaviour at high $q^2$ values
 \cite{Albertus:2014gba}. To
extend the above predictions beyond its region of applicability (near
$q^2_{\rm max}$), we adopt a multiply-subtracted Omnes dispersion
relation \cite{Albertus:2006gx, Flynn:2007ii, Flynn:2007qd, Flynn:2006vr}, 
and we take
\begin{equation}
f^+(q^2)\approx\frac{1}{M^2_{B^*}-q^2}\prod_{j=0}^n\Big[f_+(q_j^2) 
\Big(M^2_{B^*}-q^2_j\Big)\Big]^{\alpha_j(q^2)}, \alpha_j(q^2)=\prod_{j\ne k=0}^n\frac{q^2-q_k^2}{q_j^2-q_k^2}
\label{eq:omnes}
\end{equation}
for $q^2 < s_{th}=(m_{B_s}+m_K)^2$ and where the different $q_j^2 \in
]-\infty,s_{th}[$ are the different subtraction points. The values of
    the $f_+$ form factor at the subtraction points are taken as  free
    parameters that we fit to our quark model  results (valence plus $B^*$-pole) 
    at high $q^2$ and  previous light cone sum rules (LCSR) results in the low
    $q^2$ region \cite{Duplancic:2008tk}. We take the subtraction points at $q^2=0,
    q^2=\frac{q^2_{\rm max}}{3},q^2=\frac{2q^2_{\rm max}}{3}$ and
    $q^2=q^2_{\rm max}$. 
\begin{figure}[ht]
\centering
\includegraphics[width=4cm,clip]{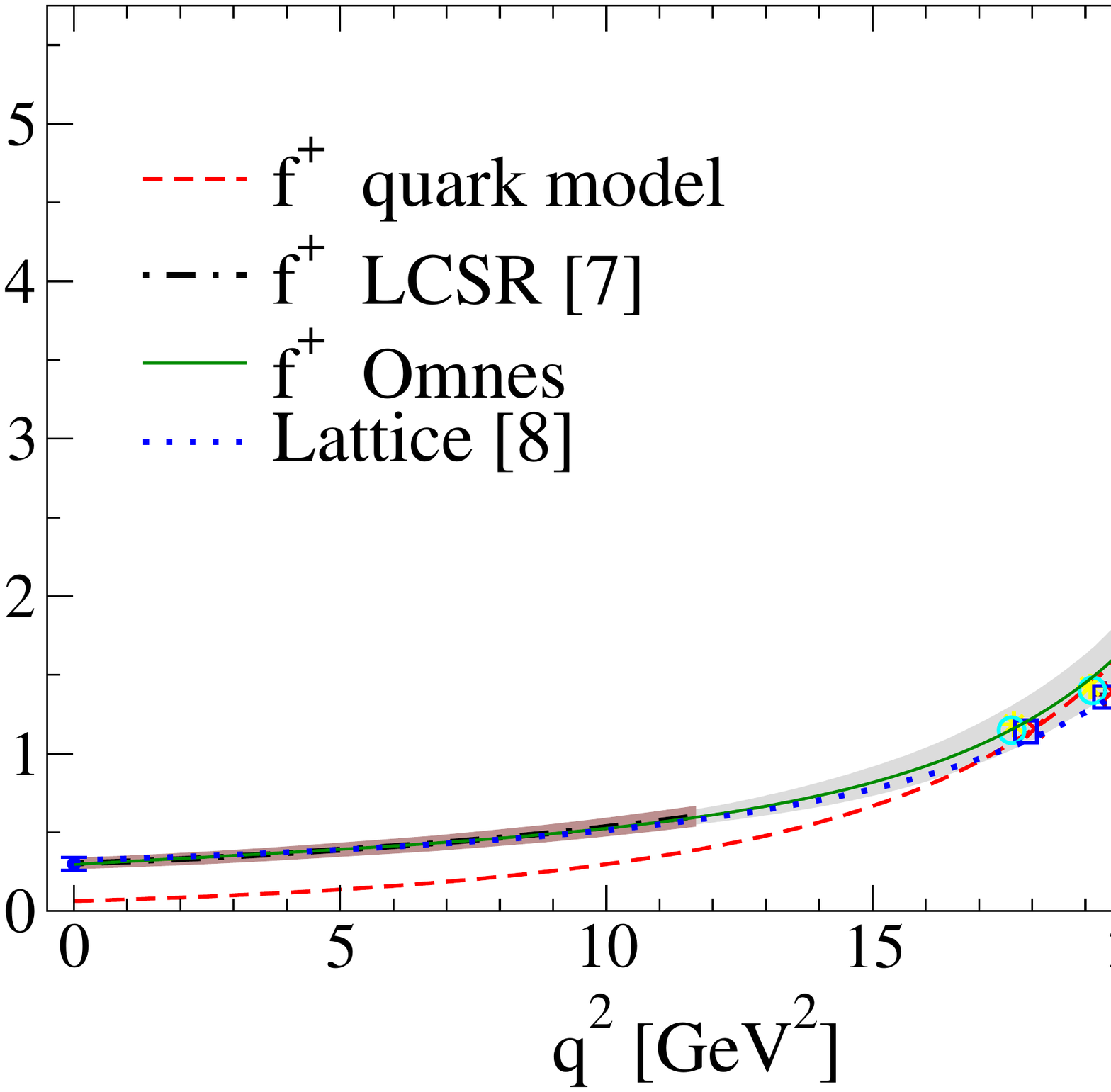}
\includegraphics[width=4.4cm,clip]{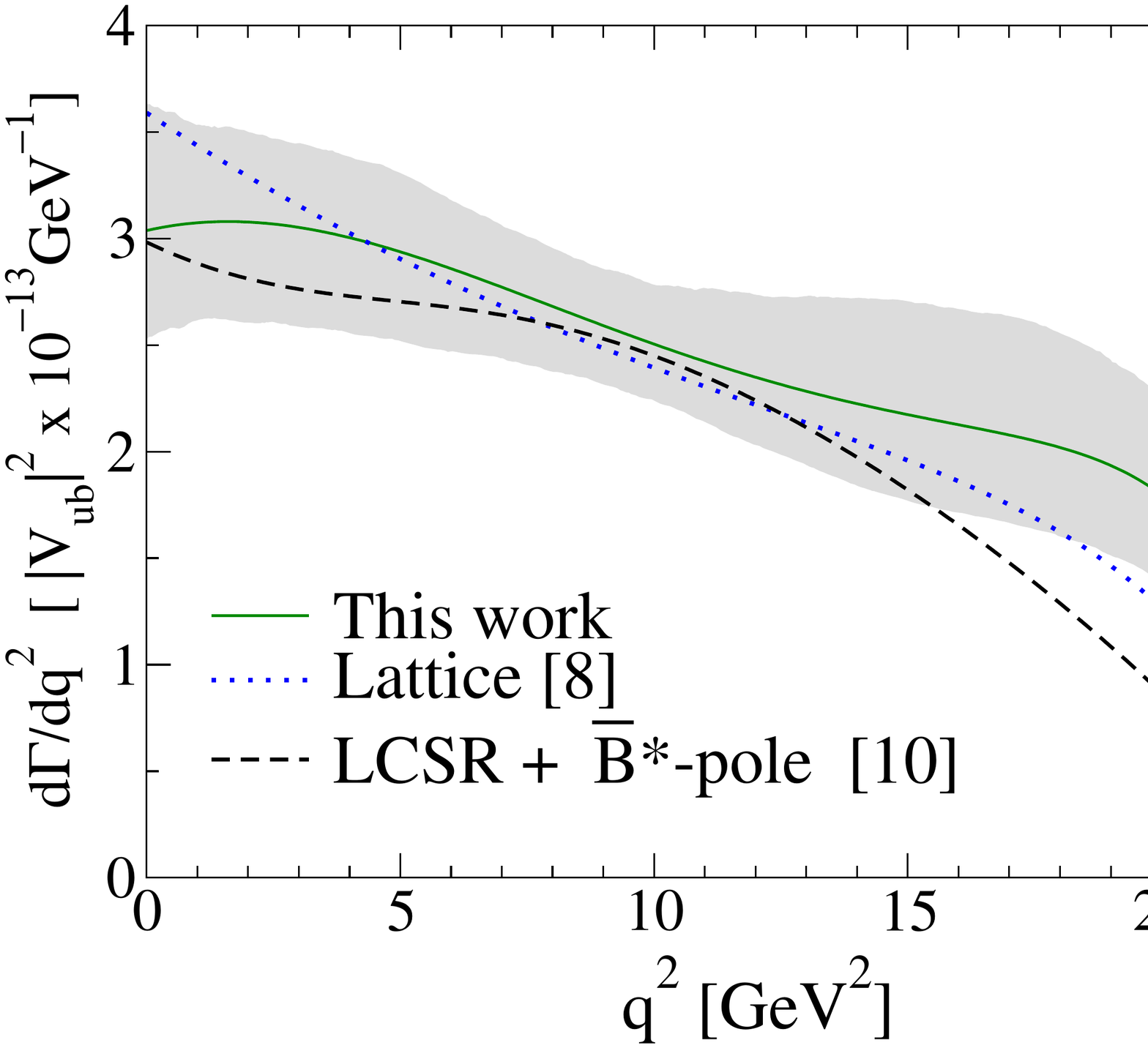}
\caption{Left panel: Comparison of the different approaches to the
  $f_+$ form factor.  Right panel: Differential decay width calculated
  with a 68\% confidence level band obtained with our fitting
  procedure}
\label{fig1}       % Give a unique label
\end{figure}
In the left panel of Fig.~\ref{fig1} we compare the $f_+$ form factor obtained
in our combined approach with the ones
calculated using LCSR techniques \cite{Duplancic:2008tk} and lattice
QCD~\cite{Bouchard:2014ypa}.
The latter are an extrapolation of the  lattice
points  obtained in Ref.~\cite{Bouchard:2013zda} which are also shown. 
   In the right panel, we plot the differential
decay width. We compare our results with the ones in Ref.~\cite{Li:2001yv},
obtained in a LCSR+$B^*$-pole calculation,  and in 
 Ref.~\cite{Bouchard:2014ypa}, obtained in lattice QCD. 
In both panels we also include a 68\%
confidence level band for our predictions.
The total decay width that we get is $
\Gamma(B_s\to Kl^+\nu_l)=(5.47^{+0.54}_{-0.46})|V_{ub}|^2\times 10^{-9}\,{\rm
MeV}$.
This is to be compared with the results
$(4.63^{+0.97}_{-0.88})|V_{ub}|^2\times
10^{-9}\,{\rm MeV}$~\cite{Li:2001yv}, where we have propagated a 10\% uncertainty 
in the
form factor, and $(5.1	\pm 1.0)|V_{ub}|^2\times
10^{-9}\,{\rm MeV}$~\cite{Bouchard:2014ypa}. The three estimates are compatible
 within uncertainties. More details are given in Ref.~\cite{Albertus:2014gba}.

\section{Semileptonic $B_s\to D^{(*)}_{s}$ decays}

We have considered the semileptonic decays of $B_s$ meson into $D^{(*)}_{s}$
states with $J^\pi$ quantum numbers $0^-$, $0^+$, $1^-$, $1^+$, $2^-$
and $2^+$. The form factor decomposition required for each channel can
be found in Ref.~\cite{Albertus:2014bfa}. We have adopted the helicity formalism of
Ref.~\cite{Ivanov:2005fd} to compute the contraction of the leptonic
and hadronic tensors. Expressions for the helicity amplitudes can be
found in Ref.~\cite{Albertus:2014bfa}

\begin{table}
\centering
\caption{Branching fractions for the indicated decay channels, in percentage.\label{tab:bra1}}
%\label{tab1}       % Give a unique label
% For LaTeX tables you can use
\begin{tabular}{c|cc}
\hline
    $M'$        &   $l=e,\mu$        &  $l=\tau$                \\\hline
%                & This work        &  This work                 \\
$D_s^+ $         &   2.32           &   0.67                     \\
$D_{s0}^{*+}  $   &   0.39           &   0.04                     \\
$D_s^{*+} $      &   6.26           &   1.53                     \\
$D_{s1}^+(2460)$ &   0.47           &   0.04                     \\
$D_{s1}^+(2536)$ &   0.32           &   0.03                     \\
$c\bar s(2^-) $ &   9.2\ten{3}     &   2.0\ten{4}               \\
$D_{s2}^{*+} $    &   0.44           &   0.03                     \\\hline
\end{tabular} 
% Or use
%\vspace*{5cm}  % with the correct table height
\end{table}

In Table~\ref{tab:bra1} we show our results for the branching ratios.
As shown in
Table V of Ref.~\cite{Albertus:2014bfa} the results of
this work are in good agreement with those from Ref.~\cite{Faustov:2012mt},
obtained in a relativistic quark model approach. The agreement is also good
with the quark model calculation of Ref.~\cite{Zhao:2006at}. Our results
for decays into orbitally excited final $D_s^*$ mesons agree with our
previous results from Ref.~\cite{segovia:2011dg}, though in that work the
potential model that has been used was much more sophisticated. Our results also compare well to the
sum-rules calculation of Refs.~\cite{Azizi:2008xy, Azizi:2008vt}, while the
result of Ref.~\cite{Blasi:1993fi} is lower by a factor of two. The same
happens if we compare with the results of Ref.~\cite{Chen:2012fi} or
Ref.~\cite{Li:2009wq}. In Ref.~\cite{Albertus:2014bfa} we also check our
results against Heavy Quark Symmetry predictions.

\section{$\bar B_s \to c\bar s M_F$}
We have also calculated the decay width for
two-meson nonleptonic reactions $\bar B_s \to c\bar s M_F$ where $M_F$ 
is a light pseudoscalar or
vector meson. These decays correspond to a $b \to c$ transition at the
quark level. These transitions are governed, neglecting penguin
operators, by the effective Hamiltonian of Eq.~(53) of
Ref~\cite{Albertus:2014bfa}. See
Refs.~\cite{Ebert:2006nz,Beneke:1999br} for further details.  We shall
work in the factorisation approximation, i. e., the hadron matrix
elements of the effective Hamiltonian are evaluated as a product of
quark-current matrix elements. One of these is the matrix element of
the $B_s$ transition to one of the final mesons, while the other
is determined by the decay constant of the other meson. In
Table~\ref{tab:brcomp2} we compare our calculations with previous
results and experimental data when available. More results are
shown in Ref.~\cite{Albertus:2014bfa}.
\begin{table}
\begin{tabular}{lccccccc}
\hline\hline
 &  This work & \cite{Faustov:2012mt} & \cite{Blasi:1993fi} & \cite{Chen:2012fi} &\cite{Li:2009wq} 
& Experiment \cite{PhysRevD.86.010001}\\\hline
$\bar B_s \to D_s^+ \pi^-$       & 0.53 & 0.35  & 0.5  & $0.27_{-0.03}^{+0.07}$   & $0.17_{-0.06}^{+0.07}$     & $0.32\pm0.4$ \\
$\bar B_s \to D_s^+ \rho^-$      & 1.26 & 0.94  & 1.3  & $0.64_{-0.11}^{+0.17}$   & $0.42_{-1.4}^{+1.7}$        & $0.74\pm0.17$\\
$\bar B_s \to D_s^+ K^-$         & 0.04 & 0.028 & 0.04 & $0.021_{-0.002}^{+0.002}$ & $0.013_{-0.004}^{+0.005}$&& \\
$\bar B_s \to D_s^+ K^{*-}$       & 0.08 & 0.047 & 0.06 & $0.038_{-0.005}^{+0.005}$ && \\
$\bar B_s \to D_{s0}^{*+} \pi^-$       & 0.10     &0.09                        &  $0.052_{-0.021}^{+0.25}$   &&&\\
$\bar B_s \to D_{s0}^{*+} \rho^-$      & 0.27     &0.22                            &  $0.013_{-0.05}^{+0.06}$   &&&\\
$\bar B_s \to D_{s0}^{*+} K^-$         & 0.009    &0.007                           &  $0.004_{-0.002}^{+0.002}$ &&&\\
$\bar B_s \to D_{s0}^{*+} K^{*-}$       & 0.16     &0.012                           &  $0.008_{-0.003}^{+0.004}$  &&&\\
$\bar B_s \to D_s^{*+} \pi^-$       & 0.45 & 0.27   & 0.2  & $0.31_{-0.02}^{+0.03}$    &                       & $0.21\pm0.06$\\
$\bar B_s \to D_s^{*+} \rho^-$      & 1.35 & 0.87   & 1.3  & $0.9_{-1.5}^{+1.5}$      &                                          & $1.03\pm2.6$ \\
$\bar B_s \to D_s^{*+} K^-$         & 0.04 & 0.021  & 0.02 & $0.024_{-0.002}^{+0.002}$                                           & \\
$\bar B_s \to D_s^{*+} K^{*-}$       & 0.08 & 0.048  & 0.06 & $0.056_{-0.007}^{+0.006}$                                           & \\
$\bar B_s \to D_{s1}^+(2460) \pi^-$   &0.15           &0.19  &                       &                        &&\\
$\bar B_s \to D_{s1}^+(2460) \rho^-$  &0.36           &0.49  &                       &                        &&\\
$\bar B_s \to D_{s1}^+(2460) K^-$     &0.012          &0.014 &                       &                        &&\\
$\bar B_s \to D_{s1}^+(2460) K^{*-}$   &0.020          &0.026 &                       &                        &&\\
$\bar B_s \to D_{s1}^+(2536) \pi^-$   &0.07          &0.029  &                       &                        &&\\
$\bar B_s \to D_{s1}^+(2536) \rho^-$  &0.19          &0.083  &                       &                        &&\\
$\bar B_s \to D_{s1}^+(2536) K^-$     &0.0054        &0.0021 &                       &                        &&\\
$\bar B_s \to D_{s1}^+(2536) K^{*-}$   &0.01          &0.0044 &                       &                        &&\\
$\bar B_s \to (2^-)^+ \pi^-$       &7.1\ten{5}    &&&&&\\
$\bar B_s \to (2^-)^+ \rho^-$      &0.0047        &&&&&\\
$\bar B_s \to (2^-)^+ K^-$         &5.2\ten{6}    &&&&&\\
$\bar B_s \to (2^-)^+ K^{*-}$       &2.2\ten{8}    &&&&&\\
$\bar B_s \to D_{s2}^{*+} \pi^-$       &0.1          &0.16   &                       &                        &&\\
$\bar B_s \to D_{s2}^{*+} \rho^-$      &0.27         &0.42   &                       &                        &&\\
$\bar B_s \to D_{s2}^{*+} K^-$         &0.008        &0.012  &                       &                        &&\\
$\bar B_s \to D_{s2}^{*+} K^{*-}$       &0.016        &0.022  &                       &                        &&\\
\hline
\end{tabular}
\caption{\label{tab:brcomp2} Branching ratios for the decays above.}
\end{table}
%\vspace{-2cm}
%
\begin{acknowledgement}
This research was supported by the Spanish Ministerio de Econom\'\i a
y Competitividad and European FEDER funds under Contracts
Nos. FPA2010-21750-C 02-02, FIS2011-28853-C02-02, and CS D2007-00042, by
Generalitat Valenciana under Contract No. PROMETEO/20090090, by Junta
de Andaluc\'\i a under Contract No. FQM-225, by the EU HadronPhysics3
project, Grant Agreement No. 283286, and by the University of Granada
start-up Project for Young Researchers contract No. PYR-2014-1. C.A.
wishes to acknowledge a CPAN postdoctoral contract and C.H.
-D. thanks the support of the JAE-CSIC Program.
\end{acknowledgement}

\bibliography{meson14}

\end{document}